\documentstyle[aps,epsfig,pre]{revtex}
\draft

\def\bea{\begin{eqnarray}}
\def\eea{\end{eqnarray}}
\def\beq{\begin{equation}}
\def\eeq{\end{equation}}

\begin{document}
\flushbottom

\wideabs{
\title{Random spreading phenomena in annealed small world networks}
\author{Jani Lahtinen$^1$, J\'anos Kert\'esz$^{1,2}$ and Kimmo Kaski$^1$}
\maketitle
\begin{center}
{\it $^1$Laboratory of Computational Engineering, Helsinki University
of Technology, P.O.Box 9400, FIN-02015 HUT, Finland}
\end{center}
\begin{center}
{\it $^2$Department of Theoretical Physics, Budapest University of
Technology and Economics, Budafoki \'ut 8, H-1111, Budapest, Hungary}
\end{center}

\maketitle

\begin{abstract}
  We study the simple random walk dynamics on an annealed version
  of a Small-World Network (SWN) consisting of $N$ nodes. This is
  done by calculating the mean number of distinct sites visited
  $S(n)$ and the return probability $P_{00}(t)$ as a function of
  the time $t$. $S(t)$ is a key quantity both from the statistical
  physics point of view and especially for characterizing the
  efficiency of the network connectedness. Our results for this
  quantity shows features similar to the SWN with quenched
  disorder, but with a crossover time that goes inversely
  proportianal to the probability $p$ of making a long range jump
  instead of being proportional to $p^{-2}$ as in quenched case. We
  have also carried out simulations on a modified annealed model
  where the crossover time goes as $p^{-2}$ due to specific time
  dependent transition probabilities and we present an approximate
  self-consistent solution to it.

PACSnumbers: 05.40.-a, 05.50.+q, 87.18.Sn
\end{abstract}
}
\section{Introduction}

During the last few years an overwhelming amount of evidence has been
accumulated about diverse networks showing ``small world'' properties.
This means that on the average an arbitrarily selected node can be 
reached from another node in very few steps despite the fact that 
only relatively small number of links are present \cite{Wattsbook}. 
Such networks include social nets \cite{soc}, the internet \cite{f3}, 
the www  of linked documents \cite{www}, scientific cooperation relations 
\cite{redner-bar} etc. Watts and Strogatz were the first to suggested 
a simple mathematical model, which reflects the small world phenomenon: 
Take a regular lattice and introduce (few) random links between any 
pairs of sites \cite {WS}.  This Small World Network (SWN) model 
interpolates between regular graphs or lattices and the so called 
Erd\H os-R\'{e}nyi random graphs \cite{Wattsbook}.

Real networks are commonly characterized by a number of parameters but
perhaps the most important measure is the average distance between
their nodes. It has turned out that there is a rich family of models
with the small world property but they differ in many other respects.
For example, the degree distribution of the nodes is Poissonian for
the SWN while many real networks are often scale free (with a power
law decay in the degree distribution) for which preferential growth
models have been introduced \cite{albsci}. Thus SWN's are very 
interesting graphs not only because of these properties but also 
because they are simple with nontrivial behavior but provide 
possibilities for exact solutions \cite{exact} and they seem to have 
a direct application in polymer physics \cite{blumen1}.

I addition to static structural properties of networks there is ever
growing interest in the dynamics processes in networks.  
In spite of this, relatively few results have sofar been published on 
the dynamics of small world models. It is expected that the underlying 
network topology should have a major impact on practically any 
phenomenon taking place on it. This is supported by the recent 
results on the the spectral density of the adjacency matrix of 
small world models, which show that these graphs produce a dramatic
deviation from the semi-circle law of random graphs \cite{spectra}.

It is our belief that spreading phenomena are perhaps the most
direct examples of dynamical processes reflecting the small world
properties. In direct spreading of e.g. a disease the nodes of the
graph get infected by the rule that infection propagates each time
step to all uninfected neighbors of already infected sites \cite{mouk}. 
Then the simplest example of non-trivial dynamics could be that of 
a diffusing particle on the SWN. This in turn is related to the
intensively studied process of random walks in random environments
\cite{havlinbook}. Recently some related papers have been published,
for example the study of spectral properties of the Laplacian on the
SWN \cite{Monasson}. While some numerical and analytic results were 
given in \cite{Pandit} for the spreading phenomenon being characterized 
by the average access time to the sites of the system, Jasch and Blumen 
\cite{Jasch} published simulation results for spreading on SWN using
random walk dynamics with the main quantity of interest being the 
average number of distinct sites vistited at a given time.

In all these studies a simple version of the SWN was considered: A
one-dimensional ring consisting of $N$ nodes with $k$-neighbor
interactions was assumed where, in addition, $N p$ new links were
introduced between arbitrarily chosen, but not yet connected nodes.
The result by Jasch and Blumen \cite {Jasch} is intuitively very
clear: whenever there is however small ratio of long range links
present in the graph, $S(t)$ changes its asymptotic behavior from
the $\sim \sqrt {t}$ relationship characteristic for the
one-dimensional case to $\sim t$ representing the long range
interconnected graph. In \cite {Jasch} a crossover formula was
presented by relating the probability of long range links $p$ to a
characteristic time $t^*$, where the crossover sets in ($p^\alpha
t^* = $const). Again, one would intuitively expect that $\alpha =2$
and that the numerically found discrepancies were occuring most
probably due to limited system sizes.  In fact we found
\cite{Lahtinen} in our numerical simulations that $\alpha =2$. This
turned out to be true not only for $S(n)$ but also for the so
called return probability $P_{00}$ \cite{Lahtinen,Scala}.

In their original form the SWN is defined with quenched disorder 
represented by the long range links. However, it is interesting to 
ask to what extent the small world properties of the dynamics depend 
on the quenched character of disorder. In order to find some answers 
we are in this paper setting to study the annealed version of the 
dynamics on SWN's, where the links are not frozen in but the walker 
has at any site the choice to make either a long range jump (with 
probability $p$) or to continue the random walk in the regular part 
of the lattice (with probability $(1-p)$). Such a model was studied 
in this context by Pandit and Amritkar \cite {Pandit}, where the 
average access time was calculated. Here we will concentrate to 
determine through analytical theory and simulations the average 
number of distinct sites visited and the return probability.

The paper is organized as follows: In the next section we introduce
the model and present the solution. In Section 3 we discuss a modified 
version of the model where the crossover time scales as $p^{-2}$ as 
in the quenched case. Finally we present a discussion and a summary.

\section{The basic model and its solution}

In order to visualize the model let us consider a ring with first 
and second neighbor connections ($k=2$), the sites of which being 
numbered from 0 to $N-1$.  In this system we let a random walker 
to start at site 0 at time $t=0$. Before the first step sets in the 
four links of site 0 are rewired by moving the other endpoint to 
any of the other sites with probability $p$. Then the walker moves to
one of the sites newly linked to its position with equal probability. 
Let us call this site $i$. After the walker arrives the site $i$ the 
original lattice is restored, the four links for $i$ are rewired as 
before, etc. In this way the small world property is maintained but 
we have got rid of the problems of quenched randomness.
 
The movement of the random walk is governed by the simple master equation:
\begin {equation}
\partial_t P_i(t)=\sum _{j=1,N}T_{ij}P_i(t) 
\end {equation} 
where the continuum time limit has been applied. Here $P_i(t)$ is the
probability that the walker is at site $i$ at time $t$ and
\begin {equation}
T_{ij}=W_{ij}-\delta_{ij} 
\end {equation} 
with $W_{ij}$ being the transition matrix, which has the following form:
\begin {equation}
  \label{Laplacian}
  {\bf W} = (1-p){\bf W}^{(S)} + p{\bf W}^{(L)}. 
\end {equation} 
Here $(S)$ and $(L)$ refer to short and long range jumps, respectively. 
The zeroth row of ${\bf W}^{(S)}$ reads as follows
\begin{equation}
W^{(S)}_{0\ \cdot} = \frac{1}{2k}(0,\underbrace{1,\ldots,1}_{k\mbox{ \small
times}},\underbrace{0,\ldots,0}_{N-2k-1\mbox{ \small
times}},\underbrace{1,\ldots,1}_{k\mbox{ \small times}}).
\end{equation}
A similar equation holds for the ${\bf W}^{(L)}$:
\begin{equation}
W^{(L)}_{0\ \cdot} = \frac{1}{N-2k-1}(\underbrace{0,\ldots,0}_{k+1\mbox{
\small times}},\underbrace{1,\ldots,1}_{N-2k-1\mbox{ \small
times}},\underbrace{0,\ldots,0}_{k\mbox{ \small times}}).
\end{equation}
The $i^{th}$ rows are then obtained by cyclically shifting the $0^{th}$ 
rows to the right. All {\bf W} and {\bf T} matrices have the Toeplitz 
form, i.e., $T_{ij}$ depends only on the difference $(i-j)$. Therefore, 
the right hand side of Eq. (1) is a convolution which leads after spatial 
Fourier transform to the following form 
\begin{equation}
  \partial _t\hat{P}_q(t) = (\hat{W}_q(t)-1)\hat{P}_q(t).
\end{equation}
With the initial condition
\begin{equation}
  P_i(0) = \delta_{0i}
\end{equation}
the formal solution is as follows 
\begin{equation}
  \label{fourier_master_equation}
    \hat{P}_q(t) = \exp[\int_0^t(\hat{W}_q(u)-1)du].
\end{equation}
This solution can be easily evaluated for the matrix {\bf W} given 
in Eqs (3)-(5).

\subsection {Return probability and spreading rate}

Let $F_{ij}(t)$ denote the probability of the random walker visiting site 
$j$ at time $t$ having started from site $i$. Then we can write
\begin{equation}
  P_{ij}(t) = \int_{0}^t F_{ij}(u)P_{jj}(t-u)du,
\end{equation}
where $P_{ij}(t)$ is the probability for the random walker to move 
from $i$ to $j$ during the time $t$. From this we get through the 
Laplace transform the following equation
\begin{equation}
  \tilde{F}_{ij}(z) = \frac{\tilde{P}_{ij}(z)}{\tilde{P}_{jj}(z)}.
\end{equation}
Now let us take $s(t)$ as the probability of observing a new site, or 
as the \emph{spreading rate} at time $t$ when the random walker started 
from site 0:
\begin{equation}
  \label{new_site}
    s(t) = \sum_{i=0}^{N-1} F_{0i}(t)
\end{equation}
By taking the \emph{return probabilities} $P_{ii}(t)$ to be the same 
for all $i$ the equation (\ref{new_site}) reads as follows
\begin{equation}
  \label{spreading_rate}
  \tilde{s}(z) = \frac{1}{\tilde{P}_{00}(z)}\sum_{i=0}^{N-1} \tilde{P}_{0i}(z)
= \frac{1}{z\tilde{P}_{00}(z)}
\end{equation}

As stated before the  quantity of interest is the average number 
of distinct sites visited, $S(t)$, which is expressed as
\begin{equation}
  \label{visited_sites}
  S(t) = \int_0^ts(u)du.
\end{equation}
Then by using Eq. (\ref{spreading_rate}) we obtain $S(t)$ by the
inverse Laplace transform of the function $\tilde {s}(z)/z$. 
It is noted that $S(t)$ could physically be intepreted as the 
amount of spreading the random walker has covered at time $t$. 

\subsection{Results}

\begin{figure}
  \centerline{\epsfxsize=8cm \epsfbox{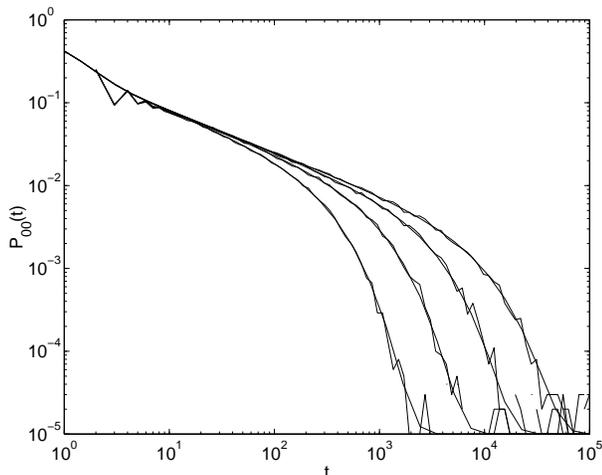}}
  \caption{The simulation results for the return probabilities for
    $p=10^{-4}$, $10^{-3.5}$, $10^{-3}$, $10^{-2.5}$ 
    (uneven line). The smooth curve shows the result of the analytical theory.}
  \label{raw_return}
\end{figure}

\begin{figure}
  \centerline{\epsfxsize=8cm \epsfbox{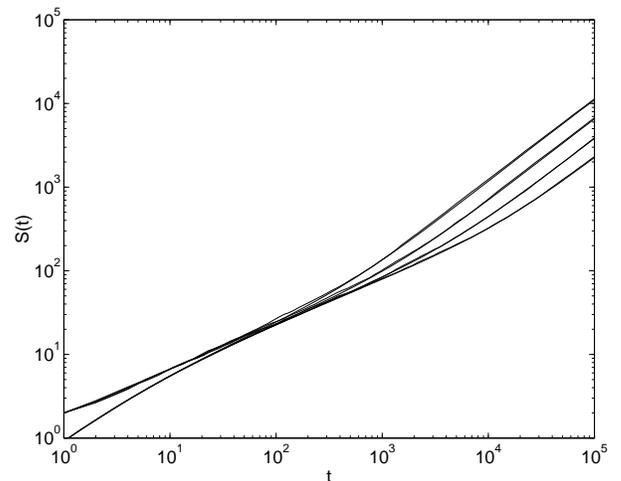}}
  \caption{The simulation results for the mean number of distinct 
   sites visited $S(t)$ (i.e. the amount of spreading) for
    $p=10^{-4}$, $10^{-3.5}$, $10^{-3}$, $10^{-2.5}$. 
   These curves start from $S(0)=1$. Analytical results, which start 
   at $S(0)=0$, are also shown.}
  \label{raw_S_I}
\end{figure}

In order to obtain the return probability and the mean number of 
visited sites as a function of the probability $p$ we have solved 
numerically the set of equations (\ref{fourier_master_equation}) 
and (\ref{spreading_rate}) given above. For the purpose of checking 
the validity of these solutions we compare them with discrete time
simulations both for $P_{00}(t)$ and $S(t)$. As depicted in 
Figs \ref{raw_return} and \ref{raw_S_I} the agreement for both 
these quantities is excellent, apart from the short time behavior, 
which is a consequence of the difference between the discrete and 
continuum time versions of the system.

\begin{figure}
  \centerline{\epsfxsize=8cm \epsfbox{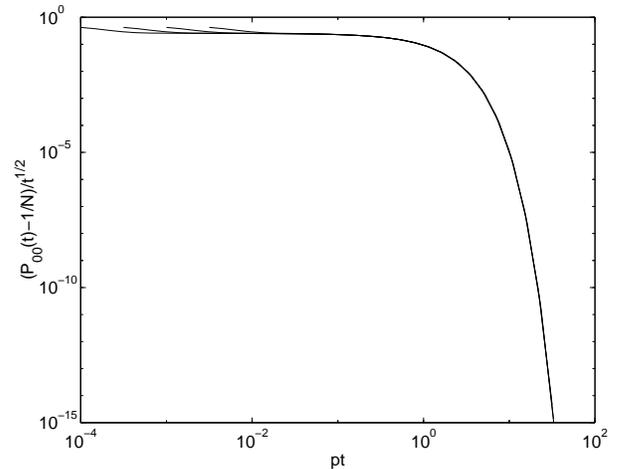}}
  \caption{The scaled return probabilities against scaled time for
    $p=10^{-4}$, $10^{-3.5}$, $10^{-3}$, $10^{-2.5}$.}
  \label{scaled_return}
\end{figure}

Fig. \ref{raw_return} shows the return probability $P_{00}$ as a
function of time. It is clearly evident from this figure that the 
simulation results agree very well with analytical calculations.  
In Fig. \ref{scaled_return} we present a scaling plot of this data, 
i.e. $(P_{00} - 1/N)\sqrt{t}$ vs. scaled time $p t$, where $1/N$ 
represents the long time limit. This scaling obeys the following 
functional form:
\begin{equation}
  \label{return_scaling}
  P_{00}(t)=t^{-1/2}\varphi(tp^\alpha)
\end{equation}
where $\varphi$ is a universal scaling function with the the property
that $ \varphi(x) \propto const $ for $x \ll 1$ and it decays rapidly
for $x > 1$.  The exponent $\alpha $ is unity (see Fig. \ref{scaled_return})
which is easily understood, because the short time characteristic 
for the one-dimensional case is valid until the walker makes a long
jump. This takes the time $1/p$, on the average, which goes as a
characteristic time into the argument of the function $f$. This is in
contrast with the behavior of the quenched model, where in order to
make a long jump, the distance proportional to $1/p$ has to be abbridged 
by diffusive motion on a one-dimensional topology of the system leading 
to a characteristic time behavior of $\sim (1/p)^2$ \cite{Jasch,Lahtinen}.

\begin{figure}
  \centerline{\epsfxsize=8cm \epsfbox{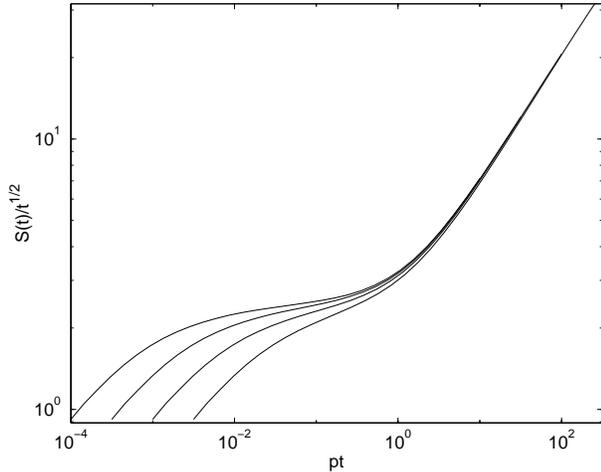}}
  \caption{The scaled spreading ($S(t)/\sqrt{t}$) against the scaled 
   time ($pt$) for
    $p=10^{-4}$, $10^{-3.5}$, $10^{-3}$, $10^{-2.5}$.}
  \label{scaled_S_I}
\end{figure}

In Fig. \ref{raw_S_I} we show the simulation and analytical results 
for the mean number of distinct sites visited, i.e. $S(t)$ as a 
function of time. Again, the good agreement between the simulation 
results and theoretical analysis is clear. The scaling form 
corresponding to (\ref{return_scaling}) would now read as
\begin{equation}
  S(t)=t^{1/2}f(tp^\alpha) 
\end  {equation} 
where the universal scaling function $f$ is of the following form
\begin {equation}
  f(x) \propto \left\{ \begin{array} {r@{\quad \mbox{for}\quad}l}
      \mbox{const} & x \ll 1 \\ \sqrt{x} & x \gg 1 
    \end{array} \right. 
\end{equation}
Here the short time behavior is $S \sim t^{1/2}$ while for long
times it becomes proportional to $t$. As seen from the scaling plot 
of Fig. \ref{scaled_S_I}, the same $\alpha=1$ applies to this case 
as for the return probability.

\section{The self--consistent model}

Next we ask the question whether it is possible to modify the annealed
model in a way that the crossover is shifted such that $\alpha  = 2$
would be obtained. This behavior is achieved by assuming a time 
dependent transition probability, as discussed below.

\begin{figure}
  \centerline{\epsfxsize=8cm \epsfbox{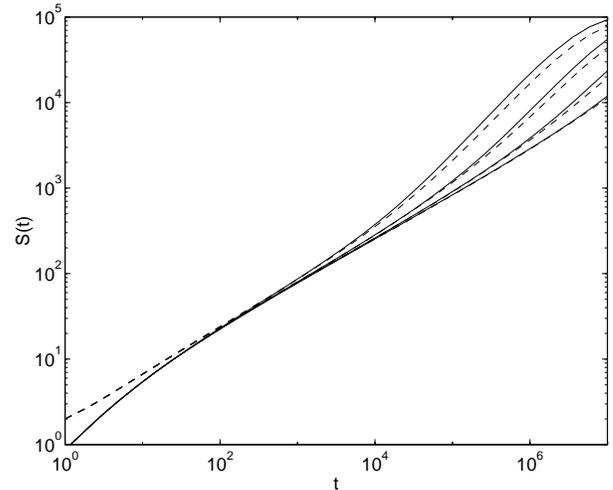}}
  \caption{The results for the spreading vs. time 
   of the self--consistent model obtained from the analytical 
   theory (solid line) and  from the quenched simulations (dashed 
   line) for $p=10^{-4}$, $10^{-3.5}$, $10^{-3}$, $10^{-2.5}$.}
  \label{model_2_S}
\end{figure}

Since the scaling of the transition occurs in the quenched system
later (as $\sim p^2t$) we replaced the multiplier $p$ of 
${\bf  W}^{(L)}$ in Eq. (\ref{Laplacian}) with $ps(t)$ to simulate 
the situation where the walker has a probability of making a long 
range leap only when visiting a previously unseen site:
\begin {equation}
  \label{self-consistent_Laplacian}
  {\bf W} = (1-p){\bf W}^{(S)} + ps(t){\bf W}^{(L)}. 
\end {equation} 
The corresponding master equation cannot be solved explicitly but we
can still estimate the solution with arbitrary accuracy by numerical
iteration scheme. In Fig. \ref{model_2_S} we show the resulting time
dependent behavior of the random walk spreading for the self-consistent 
model and simulated quenched system. Apart from the short times the 
agreement between these results seem to be quite good.   

\begin{figure}
  \centerline{\epsfxsize=8cm \epsfbox{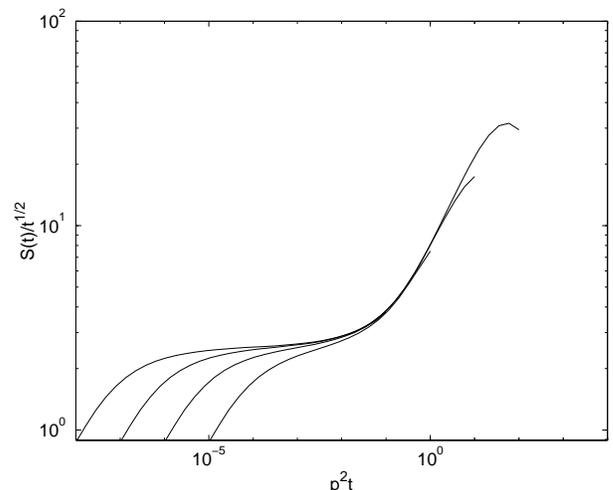}}
  \caption{The scaled spreading of the self-consistent model 
   $S(t)/\sqrt{t}$ against scaled time $p^2t$ for  
   $p=10^{-4}$, $10^{-3.5}$, $10^{-3}$, $10^{-2.5}$.}
   \label{model_2_S_scaled}
\end{figure}

In Fig. \ref{model_2_S_scaled} we show the scaling plot with 
$\alpha  = 2$ for the self-consistent model. Apart from early times
the scaling seems to hold.

Hence the numerical results justify the choice of Eq. 
(\ref{self-consistent_Laplacian}) which reflects the fact that in the
quenched model one dimensional random walk has to be carried out
between two long jumps.

\section{Summary}

We have shown that the annealed random walk model with rare long jumps
reflects some aspects of the quenched SWN's. In the simplest case,
with time independent transition probabilities, the model can be
solved. However, as expected, only qualitative agreement between the
quenched and the annealed models can be observed. With properly chosen
time dependent transition probabilities we obtain even the proper
exponent $\alpha = 2$. We think that it is quite interesting that the
behavior of the spreading can be estimated by an annealed model.

\section*{Acknowledgements} This research was partially supported by
OTKA T029985 and the Academy of Finland, project No. 1169043
(Finnish Centre of Excellence Programme 2000-2005). 

{}

\begin{thebibliography}{q}
\bibitem{Wattsbook} D.J. Watts: Small Worlds (Princeton, 1999),
S.H. Strogatz: Nature {\bf 410,} 268 (2001) R. Albert and
A.L. Baraba\'si: Rev. Mod. Phys. (in press)
\bibitem{soc} S. Milgram, Psychology Today {\bf 1}, 61 (1967)
\bibitem{www} R. Albert, H. Jeong, A. L. Barab\'asi,
Nature {\bf 401}, 130 (1999)
\bibitem {f3} M. Faloutsos, P. Faloutsos and C. Faloutsos,
Comput. Commun. Rev. {\bf 29,} 251 (1999) 
\bibitem {redner-bar} S. Rednder, Eur. Phys. J. B {\bf 4,} 131 (1998),
A.L. Barab\'asi et al. cond-mat/0104162 
\bibitem {WS} D.J. Watts, S. H. Strogatz, Nature {\bf 393}, 440
(1998)
\bibitem{albsci}  A.L. Barab\'asi, R. Albert
Science {\bf 286}, 509 (1999)
\bibitem {amral} L.A.N. Amaral et al. Proc. Natl. Acad. Sci. USA {\bf
97,} 11149 (2000)
\bibitem{exact} M.E.J. Newman, C. Moore and D.J. Watts,
Phys. Rev. Lett. {\bf 84,} 3201 (2000)
\bibitem{blumen1} S. Jespersen, I. M. Sokolov and A. Blumen,
J. Chem. Phys. {\bf 113,} 7652 (2000)
\bibitem{spectra} I.J. Farkas, I. Der\'enyi, A.L. Barab\'asi and
T. Vicsek, cond-mat/0102335
\bibitem{mouk} M.E.J. Newman and D.J. Watts, Phys. Rev. E {\bf 60}, 7332 (1999),
C.F. Moukarzel, Phys. Rev. E {\bf 60}, R6263 (1999)
\bibitem{havlinbook} B. Hughes: ``Random walks and random environments''
(Oxford UP, 1994), D. ben-Avraham and S. Havlin, ``Diffusion and
Reactions in Fractals and Disordered Systems'' (Cambridge UP, 2000)
\bibitem{Jasch} F. Jasch and A. Blumen, Phys. Rev. E {\bf 63,} 041108 (2001)
\bibitem{Lahtinen} J. Lahtinen, J. Kert\'esz and K. Kaski,
Phys. Rev. E (in press) and cond-mat/0108199
\bibitem{Pandit} S.A. Pandit and R.E. Amitkar cond-mat/00004163
\bibitem{Scala} A. Scala, L.A.N. Amaral and M. Barth\'{e}l\'{e}my,
cond-mat/0004380
\bibitem{Monasson} R. Monasson, Eur. Phys. J. B 12, 555-567 (1999)
\end{thebibliography}
\end{document}